\documentclass[pra,twocolumn]{revtex4-1}

\usepackage[a4paper,centering,hmargin=1.75cm,vmargin=2cm]{geometry}
\usepackage{amsmath,tikz,microtype,graphicx,booktabs}
\usepackage[hidelinks]{hyperref}

\begin{document}

\title{Benchmarking calculations of excitonic couplings between bacteriochlorophylls}

\author{Elise P. Kenny}
\author{Ivan Kassal}
\email{Email: i.kassal@uq.edu.au}
\affiliation{Centre for Engineered Quantum Systems, Centre for Quantum Computation and Communication Technology, and School of Mathematics and Physics, \\The University of Queensland, Brisbane QLD 4072, Australia}

\begin{abstract}
Excitonic couplings between (bacterio)chlorophyll molecules are necessary for simulating energy transport in photosynthetic complexes. Many techniques for calculating the couplings are in use, from the simple (but inaccurate) point-dipole approximation to fully quantum-chemical methods. 
We compared several approximations to determine their range of applicability, noting that the propagation of experimental uncertainties poses a fundamental limit on the achievable accuracy. In particular, the uncertainty in crystallographic coordinates yields an uncertainty of about 20\% in the calculated couplings. Because quantum-chemical corrections are smaller than 20\% in most biologically relevant cases, their considerable computational cost is rarely justified. We therefore recommend the electrostatic TrEsp method across the entire range of molecular separations and orientations because its cost is minimal and it generally agrees with quantum-chemical calculations to better than the geometric uncertainty. 
We also caution against computationally optimizing a crystal structure before calculating couplings, as it can lead to large, uncontrollable errors.
Understanding the unavoidable uncertainties can guard against striving for unrealistic precision; at the same time, detailed benchmarks can allow important qualitative questions---which do not depend on the precise values of the simulation parameters---to be addressed with greater confidence about the conclusions. 
\end{abstract}

\maketitle

Photosynthesis begins when (bacterio)chlorophyll molecules in an antenna complexes absorb light~\cite{BlankenshipBook}, creating molecular excited states. This triggers excitonic energy transfer (EET)~\cite{MayKuhn}, the migration of the excited-state energy through the network of (bacterio)chlorophyll until it decays or reaches the photosynthetic reaction center. 

The most common approach to modelling EET is to assume that each (bacterio)chlorophyll is a site---which can be in either the ground or excited states---so that the transfer of the excitation from one site to another is mediated by the coupling between them. In principle, the coupling can be calculated if the electronic structures and the relative positions of the two molecules are known. In practice, a full quantum-chemical treatment is often too expensive, which has led to various approximate methods for calculating excitonic couplings. This work is about determining the accuracy of these approximations and their range of applicability when they are applied to aggregates of bacteriochlorophylls.

At separations that are much larger than the molecular dimensions, the leading term in the excitonic coupling is the dipole-dipole interaction of the two transition dipoles. This motivates the point-dipole (PD) approximation, which truncates all the higher-order contributions. Because of its simplicity, the PD approximation has been widely used, even for calculating nearest-neighbor couplings where the small distance between the molecules would indicate that the approximation is inappropriate. 

Various methods---discussed below---go beyond the PD approximation and include more information about the electronic structure, but in a way that keeps the computation tractable. By comparing their performance with accurate quantum-chemical calculations, one might expect that a hierarchy of methods with defined distance cutoffs could be developed, so that the suitable method would be known for any particular intermolecular distance. The difficulty with that approach is that setting cutoffs depends on a subjective opinion about what is a tolerable error. 

However, an objective error threshold follows from the propagation of errors in the experimentally measured values that enter EET calculations. It is sufficient to use less accurate methods if the corrections would be drowned out by unavoidable uncertainties already in play. A particular inescapable error in EET calculations arises from the uncertainties in the atomic positions within each pigment. We show that a substantial uncertainty in the coupling is obtained by propagating the geometric uncertainties even if high-resolution crystal structures are used. In most cases, this uncertainty exceeds the quantum-mechanical corrections to the coupling, meaning that the substantially cheaper classical calculations are equally reliable.

\begin{table*}[t]
  \renewcommand{\arraystretch}{1.2}
  \begin{tabular}{p{10mm}p{42mm}p{19mm}p{28mm}p{14mm}p{24mm}}
    \toprule
    ~\newline Method & & Short-range coupling & Representation of\newline transition density & ~\newline Cost & ~\newline Benefit \\
    \midrule
    PD & Point dipole & \quad No & Point dipole & $<1\;\mathrm{ms}\,^a$ & Lowest cost \\
    ED & Extended dipole & \quad No & Two charges & $<1\;\mathrm{ms}\,^a$ & 
        \begin{tikzpicture}[overlay]
            	\draw [<->, line width=1pt, >=latex] (7ex,1.7ex) to ++(0,-11.5ex);
        \end{tikzpicture} \\
    TrEsp & \raggedright Transition charges from electrostatic potentials & \quad No & Charges on atoms & $1\;\mathrm{ms}\,^a$ & \\
    TDC & Transition density cubes & \quad No & Charges on a grid & hours$\,^{a,b}$ &  \\
    FED & Fragment excitation difference & \quad Yes & Full & 20\;h & Highest accuracy \\
    \bottomrule
  \end{tabular}
  \caption{Hierarchy of methods for calculating excitonic couplings. The cost is the typical time required to compute the coupling between two BChl molecules on a single processor. $^a\,$One-off tasks---the electronic-structure calculation of the transition density and, for TrEsp, fitting the atomic charges---are not included. $^b\,$Based on previous work~\cite{Madjet2006}.}
  \label{tbl:methods}
\end{table*}

\section{Theory of excitonic couplings}

As two molecules are brought together, the Coulomb interaction of electrons and nuclei in one molecule with those in the other increases, meaning that the eigenstates of the isolated molecules are not eigenstates of the full Hamiltonian. However, if the intermolecular interaction is weak, it is often useful to think of the system as two interacting molecules as opposed to one supermolecule. To do so, one expands the full Hamiltonian---all the interactions between particles in either molecule---in the basis of molecular states, and the off-diagonal elements of that expansion are the couplings~\cite{MayKuhn}.

Excitonic couplings are interactions between excited states localized on different molecules. In the single-exciton manifold relevant to weak illumination, the Frenkel Hamiltonian of a system of two interacting two-level molecules is
\begin{equation}
H=\begin{pmatrix}
	E_D& J \\
	J & E_A
	\end{pmatrix},
\label{eq:matrixham}
\end{equation}
where $J$ is the coupling between the donor and the acceptor, whose excitation energies are $E_D$ and $E_A$. 

The coupling is often described as containing short-range and long-range contributions. Short-range effects include exchange, overlap of donor and acceptor wavefunctions, and exciton transfer mediated by charge-transfer states~\cite{Harcourt:1994dy,Scholes:1995eg,Harcourt:1996jg,Fujimoto:2012dh}. Because they depend on the spatial overlap between donor and acceptor wavefunctions, short-range terms decrease exponentially with distance and are consequently often neglected.

Neglecting the short-range couplings leaves only the long-range Coulomb interaction, 
\begin{equation}
J_{\mathrm{Coul}}=\frac{1}{4\pi\varepsilon_0\varepsilon_r}\iint{d\mathbf{r}_{D}\,d\mathbf{r}_{A}\frac{\rho^{D}_{eg}(\mathbf{r}_{D})\rho^{A}_{eg}(\mathbf{r}_{A})}{|\mathbf{r}_{D}-\mathbf{r}_{A}|}},
\label{eq:coul}
\end{equation}
where $\mathbf{r}_{D}$ and $\mathbf{r}_{A}$ are spatial coordinates and $\rho_{eg}^X(\mathbf{r}_{X})=\varphi_{e}^X(\mathbf{r})\varphi_{g}^X(\mathbf{r})$ is the transition density between the ground ($\varphi_g^X$) and excited ($\varphi_e^X$) states of molecule $X$ (either the donor $D$ or the acceptor $A$). We return to the appropriate choice of the relative permittivity $\varepsilon_r$ below.

Because Eq.~\ref{eq:coul} resembles the interaction of two charge distributions, the integral can be simplified using elementary approximations from electrostatics. The various approximations---summarized in Table~\ref{tbl:methods}---differ in how they condense all the information in the continuous transition densities into something more manageable and discrete. 

The simplest approximation, useful for intermolecular separations much larger than the sizes of the molecules, is the point-dipole approximation (PD), obtained as the lowest-order term in the multipole expansion of Eq.~\ref{eq:coul}~\cite{MayKuhn},
\begin{equation}
J_{\mathrm{PD}}= \frac{1}{4\pi\varepsilon_0\varepsilon_r}\left(\dfrac{\mathbf{d}_{D}\cdot\mathbf{d}_{A}}{r^3}-3\dfrac{(\mathbf{d}_{D}\cdot\mathbf{r})(\mathbf{d}_{A}\cdot\mathbf{r})}{r^5}\right),
\label{eq:dipole}
\end{equation}
where $\mathbf{r}$ is the separation between the molecules (i.e., the centers of their transition densities) and $\mathbf{d}_{D}$ and $\mathbf{d}_{A}$ are their transition dipole moments, \mbox{$\mathbf{d}_{X}= \int d\mathbf{r}_X\, \mathbf{r}_X\,\rho^{X}_{eg}(\mathbf{r}_X).$}

However, PD is not accurate even at intermediate molecular separations~\cite{Howard2004,Frahmcke2006,Fuckel:2008ho,Czader:2008ht,MunozLosa:2009dl,Lee:2013fg,Kistler:2013bg}. A sequence of more accurate approximations can be obtained by representing the transition density as originating from an array of suitably chosen transition charges,
\begin{equation}
J_{\mathrm{Coul}}\approx \frac{1}{4\pi\varepsilon_0\varepsilon_r}\sum_{i \in D}\sum_{j \in A} \frac{q_i q_j}{|\mathbf{r}_i-\mathbf{r}_j|}.
\label{eq:charges}
\end{equation}
The charges $q_i$ and their positions $\mathbf{r}_i$ are chosen once and for all by fitting them to the \emph{ab initio} transition density, allowing each subsequent coupling calculation to be much faster.
The different methods that have been used---extended dipole, TrEsp, and TDC---differ only in the number of transition charges used.

In the extended dipole approximation (ED), two transition charges are used per molecule, so that the Coulomb interaction becomes
\begin{multline}
J_{\mathrm{ED}} = \frac{\delta^2}{4\pi\varepsilon_0\varepsilon_r}\left(\dfrac{1}{| \textbf{r}_D^+-\textbf{r}_A^+|} + \dfrac{1}{| \textbf{r}_D^--\textbf{r}_A^-|}- \right.\\
\left.- \dfrac{1}{| \textbf{r}_D^+-\textbf{r}_A^-|} - \dfrac{1}{| \textbf{r}_D^--\textbf{r}_A^+|}\right),
\label{eq:ext_dip}
\end{multline}
where $\mathbf{r}^\pm$ are the positions of the positive and negative charges and $\pm\delta$ is the magnitude of the charges. To be consistent with the point-dipole approximation at large separations, the charge $\delta$ and the distance $\textbf{r}_X^+-\textbf{r}_X^-$ must be chosen so that $\delta (\textbf{r}_X^+-\textbf{r}_X^-)=\mathbf{d}_X$, essentially making $\delta$ an additional free parameter, whose tunability ensures ED agrees with the exact results better than PD.

The opposite extreme is the transition density cube~(TDC) method~\cite{Krueger1998}, where the transition charges are located on a Cartesian grid, making the method a direct numerical integration of Eq.~\ref{eq:coul}. The main difficulty is that the grid needs to be fine~\cite{Madjet2006,Czader:2008ht,Maj:2012bp}, containing many charges even in areas where the transition density is negligible. For bacteriochlorophylls, the method converges when the number of charges in each molecule is around 500\,000~\cite{Madjet2006}, making the calculation of their pairwise interactions slow and prone to rounding errors. 

Between the two extremes of ED and TDC lies the transition monopole approximation (TMA)~\cite{Weiss1972,Chang1977,Sauer:1996hy}, which assigns one transition charge to each atom. If hydrogens are excluded, this leads to about 50 charges per BChl, a happy medium between 2 and 500\,000. The atomic transition charges were initially assigned using a Mulliken or Hirshfeld population analysis~\cite{Howard2004}, but this approach is not uniquely defined and was found to not accurately reproduce the shape of the transition density~\cite{Madjet2006}. The method of transition charges from electrostatic potentials (TrEsp)~\cite{Madjet2006,Renger2009} solves this problem by fitting the charges to best represent the transition density.
For molecules with few atoms, additional fitting parameters can be supplied by placing multipoles at each atom, which slightly improves the accuracy at the shortest separations~\cite{Fujimoto:2014ig,Blasiak:2015ja}.
However, 50 charges offer enough free parameters that TrEsp is as accurate as TDC for chlorophylls~\cite{Madjet2006}, which is why we do not include TDC results in this study. 

An alternative to transition charges and Eq.~\ref{eq:charges} is to expand the transition densities in a convenient chemical basis set and compute the resulting integrals in Eq.~\ref{eq:coul} using optimized quadrature techniques of quantum chemistry~\cite{Tretiak2000,Wong:2004vq,Iozzi:2004cc,Hsu2008,Fujimoto:2009fw}. Like TDC, this approach gives the exact Coulomb coupling within the chosen basis, but is more expensive than TrEsp because it still requires numerical integration every time. As with TDC, we do not consider this approach here because TrEsp is sufficiently accurate.

Once long-range couplings have been calculated using one of the methods surveyed above, any further improvement must come from including short-range effects. Short-range couplings are particularly important in polycyclic aromatic hydrocarbons (PAH)~\cite{Hsu2008,Arago:2015fr} and other flat molecules~\cite{Howard2004,Kistler:2013bg,Yamagata:2014gg} whose lack of steric hindrance allows for tight packing. Among photosynthetic complexes, large short-range couplings have been reported in LH2~\cite{Scholes1999} and in the special pair of reaction centers~\cite{Madjet2009}. We return to these cases below.

The simplest situation is the coupling between a donor and an acceptor that are identical molecules. In that case, the eigenenergies of Eq.~\ref{eq:matrixham} are \mbox{$E_{1,2} = \frac12(E_D+E_A) \pm J$}, meaning that $J$ can be obtained by halving the difference between the energies of the two excitonic states,
\begin{equation}
J=\dfrac{1}{2}(E_1-E_2),
\label{eq:Jdimer}
\end{equation}
which can be obtained from an electronic-structure calculation of the entire dimer.

The same approach can be extended to heterodimers~\cite{Scholes1999,Curutchet:2005io,Madjet2009,Hsu2008,Hsu2009}. Scholes et~al.\ used the eigenstates $E_{1,2}$ of $H$ (from the quantum-chemical treatment of the dimer) and the site energies $E_D$ and $E_A$ (from the quantum-chemical treatment of the monomers) to calculate $J$~\cite{Scholes1999}. Doing so assumed that the effective shifts in the site energies of the two molecules (induced by the presence of the other) are equal, an approximation that was removed when the method was refined by Madjet et al~\cite{Madjet2009}. They used the fact that heterodimer eigenstates are not fully delocalized, and that site-energy shifts can be computed from the extent of delocalization, which they obtained by comparing the monomer and dimer transition dipole moments. In this work, we use the closely related fragment excitation difference (FED) method~\cite{Hsu2008,Hsu2009}, which measures the delocalization in the eigenstates more directly, by determining the difference between the excitation densities on the two molecular fragments. Along with $E_{1,2}$, this suffices to calculate $J$ exactly, given a particular electronic-structure method and basis set.

\section{Results and discussion}

All coupling and energy calculations were performed on bacteriochlorophyll \textit{a} (BChl\;\textit{a}). In each case, the starting point was BChl\;\textit{a} taken from the $1\alpha$ position in the crystal structure of the LH2 complex of the purple bacterium \textit{Rhodopseudomonas acidophila}~\cite{Papiz2003}.

\subsection{Geometry optimization?}

\begin{figure}[t]
	\centering
	\includegraphics[width=8cm]{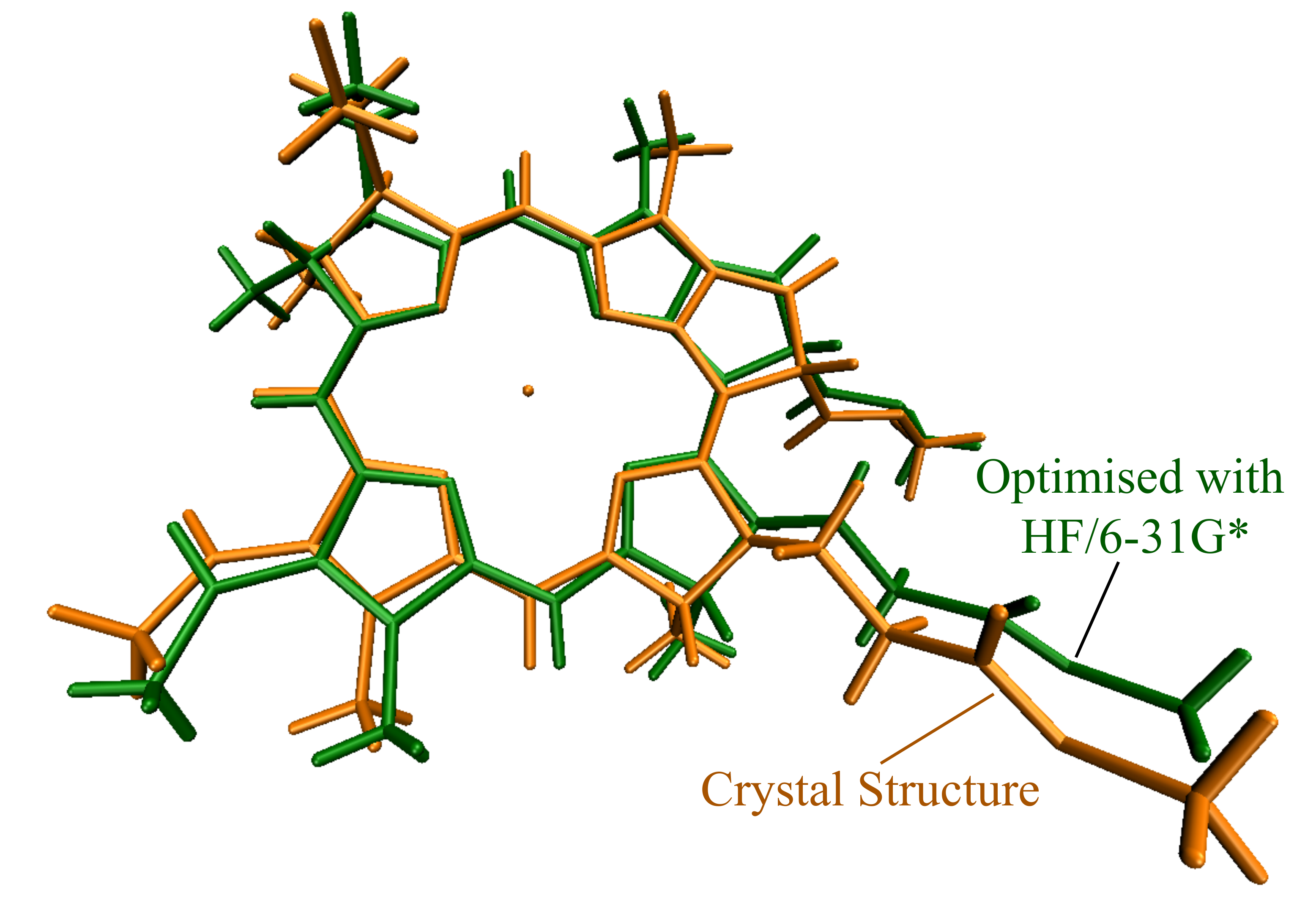}
	\caption{The crystal structure and the Hartree-Fock-optimized structure of BChl. Although the difference seems minor to the naked eye, it has a large influence on the transition energies (see Table~\ref{tbl:energies}).}
	\label{fig:optimization}
\end{figure}

An important preliminary question is which molecular geometry to use. It has been argued that the crystal structure is not reflective of the molecular configuration \textit{in vivo} and that, consequently, the geometry should be computationally optimized before calculating the excitonic couplings~\cite{Madjet2006, Madjet2009, Renger2013, Rivera:2013kb}. This optimization has generally been carried out using either HF or DFT. To the naked eye, the differences between the crystal structure and the optimized geometry appear small (Figure~\ref{fig:optimization}) and one might think that this would result in only a minor correction to the couplings.

However, the electronic properties of chlorophyll molecules are known to vary substantially with the molecular geometry~\cite{Gudowska-Nowak1990,Linnanto:2006kd,Jurinovich2015}. To test the influence of geometry optimization on BChl couplings, we compared the electronic-structure results obtained with different geometry optimizations and with experiment. We calculated the $\mathrm{Q}_y$ transition energy at the crystal structure geometry and following three different geometry optimizations: HF/\mbox{6-31G$\ast$} and DFT/B3LYP/\mbox{6-31G$\ast$} (gradient convergence criterion of \mbox{$3\cdot10^{-4} \; E_\mathrm{h}/a_0$} in each case, \mbox{Q-Chem}~4.0~\cite{QChem}), as well as molecular mechanics with Allinger's MM3 forcefield~\cite{mm3} (gradient convergence criterion of $10^{-3}$\;kcal/mol/\AA, TINKER 7.1~\cite{Tinker}). The transition energy itself was calculated using both CIS/\mbox{6-31G$\ast$} and TDDFT/B3LYP/\mbox{6-31G$\ast$}, as shown in Table~\ref{tbl:energies}. 

We found that the CIS calculation performed on the crystal structure gave the best agreement with the experimentally measured value for this transition, 770\;nm~\cite{Frigaard1996}. All geometry optimizations introduced errors much larger than the shifts in the $\mathrm{Q}_y$ peak if the spectrum is measured in different solvents or following aggregation. This suggests that geometry optimization of this molecule prior to electronic-structure calculations can introduce large and unnecessary errors. Consequently, all the following  calculations were performed at the crystal-structure geometry using CIS/\mbox{6-31G$\ast$}.

Table~\ref{tbl:energies} also compares the $\mathrm{Q}_y$ energies of the full BChl molecule and of the molecule with its phytyl tail removed. The removal of the tail changes the transition by at around 1\;nm and can therefore be safely performed in the interest of computational speed.

\subsection{Coupling calculations}

Consistency is important to ensure a fair comparison of different methods. In particular, all the underlying electronic-structure calculations must use the same method and basis set, because different methods can give substantially different results~\cite{Howard2004,Curutchet:2005io,Linnanto:2006kd,Fink:2008jv,MunozLosa:2009dl,Madjet2009,Maj:2012bp,Kistler:2013bg,Lee:2013fg,Voityuk:2014eh,Chandrasekaran:2015dc}. We used CIS/\mbox{6-31G$\ast$}, but this choice is not essential to our argument, since there is no reason to expect that other methods would be less affected by the geometric errors that are central to this paper.

Two parameters in coupling calculations are difficult to determine \textit{ab initio}: the relative permittivity $\varepsilon_r$ and the magnitude $d$ of the transition dipole moment. The vacuum value $\varepsilon_r=1$ is an appropriate choice for nearest-neighbor couplings in tightly packed aggregates, but for more distant molecules, the Coulomb interaction is screened by the intervening medium. The appropriate choice of $\varepsilon_r$ for light-harvesting environments has been discussed extensively~\cite{Renger2013}, and values from 1 to 2 have been used. This debate is beyond the scope of this work; instead, we report vacuum couplings, which can easily be adjusted by multiplication with $1/\varepsilon_r$. Our conclusions about the sizes of relative errors are unaffected, because all the couplings are scaled in the same proportion.

The second uncertain parameter is the magnitude of the $\mathrm{Q}_y$ transition dipole moment, which tends to be overestimated by CIS calculations~\cite{Krueger1998,Scholes1999,Madjet2006,Renger2009}. For the crystal-structure geometry, CIS/\mbox{6-31G$\ast$} predicts a $\mathrm{Q}_y$ dipole moment of $d_\mathrm{CIS}=10.45$\,D, significantly larger than the best experimental estimate of 6.1\,D (in vacuum)~\cite{Knox2003}. Here again we make the simplest choice, reporting all results with the theoretically predicted value $d_\mathrm{CIS}$, which can be corrected by multiplication with $(d/d_\mathrm{CIS})^2$, where $d$ is the desired magnitude of the dipole moment~\cite{Krueger1998,Scholes1999,Madjet2006,Renger2009}. Again, the conclusions about the relative errors are unaffected because all couplings would be scaled by the same factor.

\begin{figure}[t]
\centering
\includegraphics{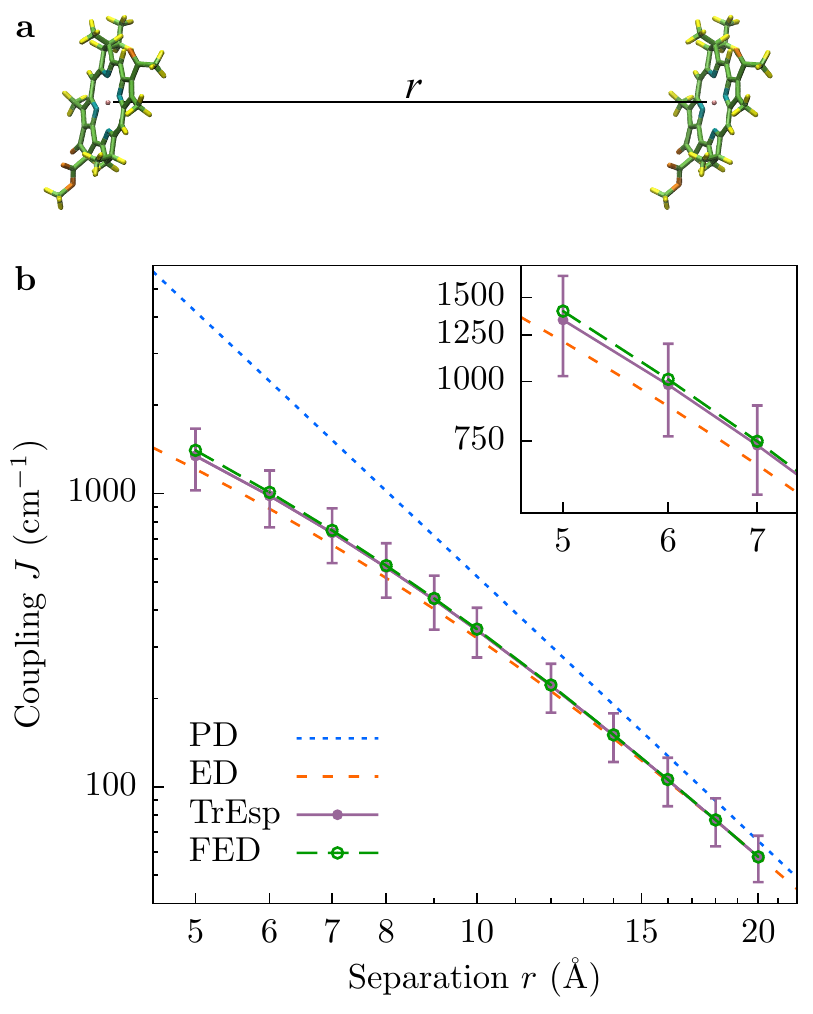}
\caption{Excitonic couplings for the $\mathrm{Q}_y$ transition of  in vacuum. 
\textbf{(a)} The geometry was taken from the crystal structure of LH2 of \textit{Rps.~acidophila}~\cite{Papiz2003}, with the second molecule a copy of the first, displaced perpendicular to the bacteriochlorin ring.
\textbf{(b)} Comparison of four different methods (log-log scale). The error bars on the TrEsp values indicate the uncertainty in the coupling (20\% on average), propagated from the uncertainty in the crystal-structure atom coordinates. Since the uncertainty is much larger than the difference between TrEsp and the fully quantum FED, the additional computational cost of FED is not justified. 
\textit{Inset:} The deviation of TrEsp and ED from FED at short distances.
} \label{fig:separation}
\end{figure}

Having made these preliminary choices, we compared the couplings predicted by PD, ED, TrEsp, and FED, calculated between two identical BChl molecules, displaced perpendicular to their bacteriochlorin rings by a separation ranging from 5\,\AA{} to 20\,\AA{} (Figure~\ref{fig:separation}a).

\begin{table}[t]
	\centering
	\begin{tabular}{lcccc}
            \toprule
             & \multicolumn{4}{c}{$\mathrm{Q}_y$ transition (nm)} \\ 
             \cmidrule(lr){2-5} 
             & \multicolumn{2}{c}{With tail} & \multicolumn{2}{c}{Tail removed} \\ 
             \cmidrule(lr){2-3} \cmidrule(lr){4-5} 
            Geometry  & CIS & TDDFT & CIS & TDDFT \\
            \midrule
            Crystal structure   & 794 & 621 & 793  & 621 \\
            Optimized: & & & & \\
            \quad HF/6-31G* & 341 & 559 & 341 & 559 \\
            \quad B3LYP/6-31G* \quad & 664 & 568 &  667 & 568 \\
            \quad MM3 & 611 & 578 &  610 & 578 \\
            \midrule
            Experiment~\cite{Frigaard1996} & \multicolumn{4}{c}{770} \\
            \bottomrule
	\end{tabular}
        	\caption{The influence of geometry on the calculated wavelength of the $\mathrm{Q}_y$ transition of BChl\;\textit{a}. The transition was calculated using both CIS and TDDFT based on the crystal structure and on the geometries obtained by optimizing the crystal structure using different routines. The best agreement with experiment is obtained at the crystal structure geometry, indicating that geometry optimization may induce large and uncontrolled errors. In addition, removing the phytyl tail has a negligible effect, as expected.}
	\label{tbl:energies}
\end{table}

For PD, the magnitude and direction of the transition dipole moment were obtained from the CIS calculation. 
For ED, the two transition charges were chosen so that the ED coupling equaled the FED coupling at 20\,\AA{}, giving a dipole length of 10.2\,\AA. The dipole length is sensitive to the geometries used for the fitting, which is why our value differs from Renger's 8.8\,\AA~\cite{Renger2009}.
For TrEsp, the best accuracy would be obtained by recalculating the transition charges at every geometry, but that would require an electronic-structure calculation and electrostatic fitting each time, defeating TrEsp's purpose as a fast method. Here, we use the original transition charges that the authors of TrEsp recommended be used even if the molecule undergoes slight configurational change~\cite{Madjet2006}. However, because those charges were calculated for a planar BChl\;\textit{a} and predict a transition dipole of 10.11\,D, we scaled all the charges by $d_\mathrm{CIS}/10.11\,\mathrm{D}$ to ensure a consistent comparison with the other methods.
For FED, we employed the routine implemented in Q-Chem~4.0~\cite{QChem}. 

Figure~\ref{fig:separation}b shows that all the coupling methods converge to a common value at large separations, as expected. At the smallest separation in Figure~\ref{fig:separation}b, the FED couplings differ from the TrEsp couplings by only~3\%, indicating that the short-range contribution, ignored by TrEsp, is small compared to $J_\mathrm{Coul}$. 

For separations under 5\,\AA, FED results became difficult to interpret because the molecules are so strongly coupled that it becomes impossible to speak of two coupled molecules and one must treat the dimer as a supermolecule. In particular, if the coupling becomes comparable to the spacing between electronic excited states, higher-energy transitions will contaminate the calculation and the two-state model of coupled $\mathrm{Q}_y$ transitions will fail. The breakdown of the two-state approximation is easily diagnosed in the parallel homodimer, because the approximation predicts that one of the two dimer states will be perfectly bright (twice the oscillator strength of the monomer) and the other perfectly dark (zero oscillator strength). For parallel BChls, this condition fails to hold around 4\,\AA, making it dangerous to calculate couplings by simply halving the energetic gap. 

An Mg--Mg separation of 5\,\AA{} is small and can only occur in configurations close to parallel because of the size of the BChl molecules. For comparison, the most strongly coupled naturally occurring BChls (with a known crystal structure) are the special pairs of reaction centers, with a Mg--Mg distance of about 8\,\AA{}~\cite{Stowell:1997ja}. FED can fail in some of these cases as well, as we discuss below.

\subsection{Uncertainties}

Most theoretical calculations include experimentally measured quantities at some point and are therefore subject to the propagation of errors. In particular, crystal-structure atomic coordinates carry uncertainties due to thermal motion and the limited resolution of the instrument. Those uncertainties should be propagated through the entire calculation.

Uncertainties in crystallographic coordinates are specified using Debye--Waller factors, also known as temperature factors or $B$ factors~\cite{Willis1975}. They are given for each atom in the standard PDB crystallographic format, and are proportional to the mean-squared fluctuations of atomic positions, 
\begin{equation}
B=\frac{8\pi^2}{3}\langle u^2\rangle.
\end{equation}
Hence, the standard uncertainty of each Cartesian coordinate is 
$\sigma_i=\sqrt{\langle u^2\rangle/3}=\sqrt{B/8\pi^2}$. 
For the BChl molecule from the crystal structure of LH2~\cite{Papiz2003}, the average $B$ factor is $16\;\mathrm{\AA}^2$, corresponding to an uncertainty of 0.45\;\AA{} in each Cartesian coordinate.

We propagated these geometric uncertainties through our TrEsp calculations, and the results are shown as error bars in Figure \ref{fig:separation}b. For each calculation, the atomic positions in both molecules were chosen randomly from a normal distribution centered at the atom's reported coordinates and with the standard deviation determined from the $B$ factor. This was repeated with a sample of 1000 random geometries to give the distribution of TrEsp couplings.
We found that, for the parallel geometry, the resulting uncertainty in the TrEsp couplings (one standard deviation) was about 20\% across the whole range of separations. Since this is larger than the difference between the TrEsp and the FED values, we conclude that the considerable computational cost of FED is not justified at any separation.

Our results are consistent with those of Arag\'o and Troisi, who used molecular dynamics simulations of anthracene crystals to find that the thermal nuclear motion caused large fluctuations in excitonic couplings~\cite{Arago:2015fr}. Although their results were dominated by short-range couplings---which are substantial in PAHs---our results confirm the importance of geometric fluctuations even for long-range couplings.

The errors bars in Figure \ref{fig:separation}b are only the lower bound on the uncertainty in the couplings both because realistic $B$ factors are probably larger and because we did not consider other sources of uncertainty. We used the $B$ factors from the LH2 crystal structure taken at~100\,K; at physiological temperature, the thermal motion of the atoms would be greater. In addition, some simulations indicate that $B$ factors computed in the course of ordinary structure refinement may underestimate the magnitude of thermal fluctuations~\cite{Kuzmanic2014}. The other kinds of uncertainty that may lead to substantially larger error bars include uncertainties in the choice of the transition dipole moment or the relative permittivity (especially at small-to-intermediate separations where there is little medium between the molecules~\cite{Curutchet:2007ji}). 
Because these additional uncertainties could only increase the error bars, including them would strengthen our argument that the agreement between TrEsp and FED is better than the unavoidable error in the calculations.

\begin{figure}[t]
\centering
\includegraphics{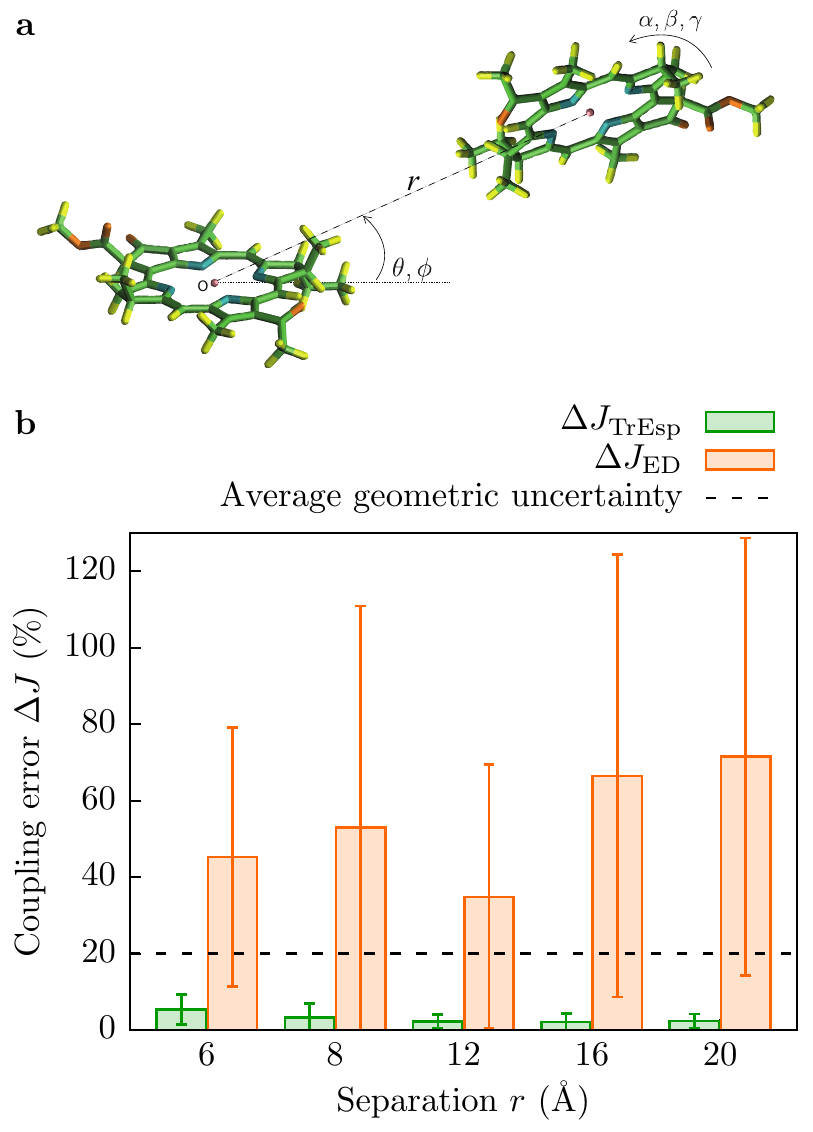}
\caption{Accuracy of the TrEsp and ED methods. 
\textbf{(a)} The separation $r$ between the magnesium atoms was fixed and the position (spherical coordinates $r$, $\theta$, $\phi$) and orientation (Euler angles $\alpha$, $\beta$, $\gamma$) of the second molecule were randomized (cases with overlapping molecules were rejected).
\textbf{(b)} Error of TrEsp and ED, with respect to FED, averaged over 50 random orientations at each separation. The average error is shown, together with error bars at one standard deviation. The dashed black line is the typical error arising from the uncertainty in the molecular geometry (the error bars in Figure \ref{fig:separation}b). 
ED can lead to large errors, indicating that the relatively good agreement in Figure~\ref{fig:separation}b was coincidental. By contrast, TrEsp performs better than the propagated geometric uncertainty in almost all cases, confirming it as a robust method. 
}\label{fig:orients}
\end{figure}

\subsection{Other orientations}

To ensure that the good agreement between TrEsp and FED in Figure~\ref{fig:separation}b was not peculiar to the parallel arrangement of the two BChls, we repeated the calculation at different relative positions and orientations of the two molecules. We chose five molecular separations (\mbox{Mg--Mg} distance) and calculated the ED, TrEsp, and FED couplings for 50 random orientations at each separation (see Figure~\ref{fig:orients}b). Relative orientations were excluded if the two molecules collided, i.e., if any atom from the first molecule was closer than 1.5\,\AA{} from any atom in the second.

To determine the accuracy of the two approximate methods, the error of the ED and TrEsp calculations, with respect to the FED couplings, was calculated at each orientation. The typical relative error at each molecular separation was defined as
\begin{equation}
\Delta J_\mathrm{ED/TrEsp}=\frac{\langle\, |J_\mathrm{FED}-J_\mathrm{ED/TrEsp}|\,\rangle}{\langle J_\mathrm{FED} \rangle},
\end{equation}
where the averaging $\left<\cdot\right>$ was carried out over all the relative orientations at that molecular separation. This error is shown in Figure~\ref{fig:orients}b, together with the standard deviation of the errors, $|J_\mathrm{FED}-J_\mathrm{ED/TrEsp}|/\left<J_\mathrm{FED}\right>$.

The typical error of the ED approximation was large---about 50\% of the FED coupling---indicating that the method is not reliable in general and that the apparently good agreement in Figure~\ref{fig:separation}b occurred because the ED dipole length was fitted to the FED data for the parallel arrangement. 

By contrast, TrEsp maintained the accuracy from Figure \ref{fig:separation}b across the random orientational ensemble, with an average error of less than 6\% at each separation. For all separations, the geometric uncertainty is more than one standard deviation higher than the average error of TrEsp. This indicates that it is unnecessary to use FED to calculate couplings between BChl molecules, as the limit of accuracy is not the choice between TrEsp and FED, but the uncertainty in the atomic positions.

\begin{table}[t]
  \begin{tabular}{lcrcllll}
    \toprule
     & & \multicolumn{3}{c}{Coupling (cm$^{-1}$)} & \multicolumn{2}{c}{Relative error}  \\
          \cmidrule(lr){3-5} \cmidrule(lr){6-7} 
     Pair & $r$ (\AA) & $J_\mathrm{ED}$ & $J_\mathrm{TrEsp}$ & \!$J_\mathrm{FED}$ & $~\Delta J_\mathrm{ED}$ & $\Delta J_\mathrm{TrEsp}$\\
    \midrule
    LH2 & & & & & & \\
    \quad$J_{1\alpha1\beta}$ & 9.0 & 1099 & $704 \pm 326$ & 814 & $+35\%$ & $-14\%$ \\
    \quad$J_{1\beta2\alpha}$ & 9.0 & 214 & $592 \pm 238$ & 781 & $-73\%$ & $-24\%$ \\
    RC & & & & & & \\
    \quad$J_\mathrm{P_1P_2}$ & 7.6 & 35 & $363 \pm 371$ & 643$\,^a$ & $-95\%\,^a$  & $-44\%\,^a$ \\
    \bottomrule
  \end{tabular}
  \caption{Comparison of ED, TrEsp and FED for some of the most strongly coupled bacteriochlorophylls in photosynthetic complexes. The couplings were calculated, based on the crystal structures, for the nearest-neighbors in the B850 subunit of the LH2 complex of \textit{Rhodopseudomonas acidophila}~\cite{Papiz2003} and for the special pair in the reaction center of \textit{Rhodobacter sphaeroides}~\cite{Stowell:1997ja}. As above, we assumed the molecules to be in vacuum and, for each molecule, we assumed transition dipoles predicted by CIS; therefore, the couplings should be scaled for comparison with previous work. $r$ is the Mg--Mg separation and TrEsp values also include the geometric uncertainty. $^a\,$The FED calculation is unreliable due to the failure of the two-level approximation (see text). 
  }
  \label{tab:closest}
\end{table}

\subsection{Most difficult cases}

The means and standard deviations in Figure~\ref{fig:orients}b indicate that the difference between TrEsp and FED is less than the geometric uncertainty in the vast majority of cases. However, focusing on the mean and standard deviation risks ignoring the outliers, which may indicate additional failures of TrEsp. Of particular concern are configurations where the two BChls are parallel but offset from each other. The offset can give rise to a relatively large Mg--Mg distance even though portions of the two molecules might be close to each other. The offset-parallel arrangement arises in some natural complexes, giving rise to some of the most strongly coupled BChls in nature. To check the applicability of TrEsp to those cases, we calculated the couplings between three such pairs, as shown in Table \ref{tab:closest}.

For the two nearest-neighbor couplings in the LH2 complex of purple bacteria, the error of ED with respect to FED is over 50\% on average, confirming its poor performance seen in Figure~\ref{fig:orients}b. The error of TrEsp is, as expected, larger than the average in Figure~\ref{fig:orients}b, at 14\% and 24\% for $J_{1\alpha1\beta}$ and $J_{1\beta2\alpha}$, respectively, a result consistent with previous work finding short-range corrections of 17\% and 24\% in the two cases~\cite{Scholes1999}. However, the geometric errors are also larger, because small displacements of particular atoms can have an outsized influence on the coupling when those atoms are close together. In the two cases from LH2, the average geometric uncertainty in the TrEsp couplings is 43\%, substantially more than the 19\% error with respect to FED. Thus, we can again conclude that the uncertainty in the atomic coordinates is a larger source of error than excluding the short-range contributions.

Special pairs in reaction centers are even more strongly coupled, and the results in Table~\ref{tab:closest} indicate that all the methods fail. In particular, FED shows a considerable admixture of higher excited states, indicating that it is inappropriate to model this case as two coupled Q$_y$ transitions (Eq.~\ref{eq:matrixham}). Madjet et~al.~\cite{Madjet2009} checked the consistency of the same calculation by comparing the values of the $Y_a$ and $Y_b$ parameters of their theory; these values were unequal by a large margin, especially for certain electronic-structure methods, indicating the failure of the effective two-state Hamiltonian. The difference between our coupling and that of Madjet et~al. is probably caused by their optimization of the geometry. The failure of FED indicates that the special pair cannot be considered as two coupled molecules but should be seen as one unit. This can also be diagnosed from the severe failure of TrEsp, whose geometric uncertainty is over~100\%. 

Therefore, the geometric uncertainty indicates not only the inherent uncertainty of TrEsp, but also the applicability of the two-state approximation. In difficult cases such as those just discussed, the close separation between certain atoms will increase the uncertainty above the typical 20\%, indicating that the calculation should be viewed with suspicion.

\section{Conclusions}

We have presented a detailed comparison of methods for excitonic coupling calculations of bacteriochlorophylls, with a focus on the uncertanties that arise due to the uncertainties in the atomic positions. For TrEsp, these geometric uncertainties are much larger than the disagreement between TrEsp and the more accurate FED, indicating that the short-range contributions to the couplings are almost always a minor correction in comparison with the error bars, making the computational cost of obtaining them (6--7 orders of magnitude more than for TrEsp) unjustified.
Furthermore, TrEsp is clearly preferrable over other the PD and ED approximations, both of which lead to errors much larger than the geometric uncertainty.
Therefore, we recommend the use of TrEsp for the calculation of excitonic couplings between bacteriochlorophylls at all separations and orientations, with the warning that particularly large geometric uncertainties may indicate a failure of the two-state approximation.

Our error analysis can be extended to site energies and other components of EET simulations in order to determine the overall sensitivity to uncertainties in the experimental data. This will allow us to identify the theoretical predictions that do not depend sensitively on microscopic details and are thus more likely to apply to a wide range of pigment-protein complexes.

\section*{Acknowledgments}
This work was supported by the Australian Research Council through a Discovery Early Career Researcher Award (DE140100433) and the Centres of Excellence for Engineered Quantum Systems (CE110001013) and Quantum Computation and Communication Technology (CE110001027). The molecular structures were drawn using VMD~\cite{VMD}.

\end{document}